\documentclass[twocolumn,aps,prl,final,superscriptaddress,showpacs]{revtex4-1}
\usepackage[latin9]{inputenc}
\usepackage[active]{srcltx}
\usepackage{color}
\usepackage{amsmath}
\usepackage{graphicx}
\usepackage[unicode=true,pdfusetitle,
 bookmarks=true,bookmarksnumbered=true,bookmarksopen=true,bookmarksopenlevel=1,
 breaklinks=true,pdfborder={0 0 1},backref=false,colorlinks=true]
 {hyperref}
\hypersetup{
 linkcolor=blue,urlcolor=blue,citecolor=blue,pdfstartview={FitH},hyperfootnotes=false}

\makeatletter
\usepackage{dcolumn}% Align table columns on decimal point
\usepackage{bm}% bold math
\usepackage{times}

\makeatother

\begin{document}
\title{Energy storage in lead-free Ba(Zr, Ti)O$_{\textrm{3}}$ relaxor ferroelectrics:
\\
 Large densities and efficiencies and their origins}
\author{Zhijun Jiang}
\affiliation{MOE Key Laboratory for Nonequilibrium Synthesis and Modulation of
Condensed Matter, Shaanxi Province Key Laboratory of Advanced Functional
Materials and Mesoscopic Physics, School of Physics, Xi'an Jiaotong
University, Xi'an 710049, China}
\affiliation{Physics Department and Institute for Nanoscience and Engineering,
University of Arkansas, Fayetteville, Arkansas 72701, USA}
\author{Sergey Prosandeev}
\affiliation{Physics Department and Institute for Nanoscience and Engineering,
University of Arkansas, Fayetteville, Arkansas 72701, USA}
\author{L. Bellaiche}
\affiliation{Physics Department and Institute for Nanoscience and Engineering,
University of Arkansas, Fayetteville, Arkansas 72701, USA}
\begin{abstract}
An atomistic first-principles-based effective Hamiltonian is used
to investigate energy storage in Ba(Zr$_{0.5}$Ti$_{0.5}$)O$_{3}$
relaxor ferroelectrics, both in their bulk and epitaxial films' forms,
for electric fields applied along different crystallographic directions.
We find that the energy density linearly increases with temperature
for electric fields applied along the pseudocubic {[}001{]}, {[}110{]}
and {[}111{]} directions in Ba(Zr$_{0.5}$Ti$_{0.5}$)O$_{3}$ bulk.
For films at room temperature, the energy density adopts different
behaviors (i.e., increase \textit{versus} decrease) with strain depending
on the direction of the applied electric fields. We also predicted
ultrahigh energy densities (basically larger than 100 J/cm$^{3}$)
with an ideal efficiency of 100\% in all these Ba(Zr$_{0.5}$Ti$_{0.5}$)O$_{3}$
systems. In addition, a phenomenological model is used to reveal the
origin of all the aforementioned features, and should be applicable
to other relaxor ferroelectrics. 
\end{abstract}
\maketitle

\section{Introduction}

Dielectric capacitors are particularly promising for high-power energy
storage applications because of their ultrafast charging/discharging
rates and high reliability \cite{Chu2006,Li2015,Peng2015,Prateek2016,Palneedi2018,Yang2019,Pan2019}.
However, dielectric capacitors have relatively low energy densities
and efficiencies, which is the main bottleneck towards applications
in electronics and electric power systems. Intensive efforts have
been devoted to find novel materials with higher energy density and
efficiency. Recently, a special class of ferroelectrics with a frequency-dependent
dielectric response-versus-temperature and several characteristic
temperatures \cite{Cross1994,Bokov2006,Shvartsman2012,Akbarzadeh2012,Prosandeev2015},
namely the relaxor ferroelectrics, have been attracting much attention
because of their ultrahigh energy densities and efficiencies \cite{Instan2017,Kim2020,Pan2021}.
One can for instance cite Ba(Zr$_{x}$Ti$_{1-x}$)O$_{3}$ thin films
with a recoverable energy density of 158 J/cm$^{3}$ and an efficiency
of 72.8\% \cite{Instan2017}, 0.68Pb(Mg$_{1/3}$Nb$_{2/3}$)O$_{3}$-0.32PbTiO$_{3}$
(PMN-PT) films with 133 J/cm$^{\textrm{3}}$ and 75\% \cite{Kim2020},
and Sm-doped $y$BFO-(1$-$$y$)BTO (Sm-BFBT) with 152 J/cm$^{\textrm{3}}$
and a marked enhancement of efficiency above 90\% at an electric field
of 3.5 MV/cm \cite{Pan2021}. It is also worth noting that relaxor
ferroelectrics in the range between the Burns temperature $T_{\textrm{b}}$
and temperature $T_{\textrm{m}}$ at which the dielectric constant
exhibits a peak\textcolor{black}{{} can be defined as} superparaelectric
relaxor ferroelectrics \cite{Pan2021,Cross1987}, which is highly
promising for energy storage applications. As a matter of fact, the
superparaelectric state in relaxor ferroelectrics can exhibit large
polarizability at high electric fields with nonlinear $P$-$E$ loop
behavior while maintaining very small hysteresis \cite{Glazounov1995}.

Despite these impressive progresses on the use of relaxor ferroelectrics
for energy storage, several questions remain unaddressed, to the best
of our knowledge. For instance, can \textit{ab-initio-based methods}
reproduce the experimental finding about ultrahigh energy density
in the lead-free relaxor Ba(Zr$_{x}$Ti$_{1-x}$)O$_{3}$ system,
and, if yes, can they provide new insight into its origin? The effect
of the direction of the applied electric field on energy storage density
and efficiency is also presently unknown in this material, to the
best of our knowledge. Similarly, we are not aware that the consequence
of the epitaxial strain on energy-storage-related properties has ever
been investigated, and thus revealed and explained, in films made
of Ba(Zr$_{x}$Ti$_{1-x}$)O$_{3}$.

The aim of this article is to answer all these questions, by using
an atomistic first-principles-based effective Hamiltonian and analyzing
its results via a simple and straightforward phenomenological model.
In particular, we demonstrate, and explain why, ultrahigh energy density
and efficiency can be achieved in Ba(Zr$_{0.5}$Ti$_{0.5}$)O$_{3}$
ferroelectric relaxors, both in their bulk and epitaxial films' forms.
The effects of the direction of the applied field as well as epitaxial
strain are also revealed and understood.

This article is organized as follows. Section II provides details
about the atomistic effective Hamiltonian method used here. Sections
III A and III B report and explain energy-storage results in Ba(Zr$_{0.5}$Ti$_{0.5}$)O$_{3}$
bulk and films, respectively. Finally, Section IV concludes this work.

\section{Methods}

Here, the first-principles-based effective Hamiltonian ($H_{\textrm{eff}}$)
approach that has been developed and used in Refs. \cite{Akbarzadeh2012,Prosandeev2013,Prosandeev2013_film,Jiang2017}
is employed to investigate bulk and epitaxial films made of Ba(Zr$_{0.5}$Ti$_{0.5}$)O$_{3}$
(BZT) solid solutions. This $H_{\textrm{eff}}$ successfully reproduced
(i) the existence of temperatures characteristic of relaxor ferroelectrics
\cite{Akbarzadeh2012} (such as the Burns temperature $T_{\textrm{b}}$
that typically marks the existence of dynamical polar nanoregions
(PNRs) \cite{Akbarzadeh2012,Burns1983}, $T_{\textrm{b}}$ $\simeq$
450 K, the so-called $T^{*}$ temperature at which static PNRs typically
appear \cite{Akbarzadeh2012,Dkhil2009,Svitelskiy2005},\textbf{ $T^{*}$}
$\simeq$ 240 K, and $T_{\textrm{m}}$ $\simeq$ 130 K, at which the
dielectric response adopts a peak \cite{Cross1994,Akbarzadeh2012});
(ii) polar nanoregions \cite{Akbarzadeh2012,Prosandeev2013}; and
(iii) the unusual dielectric relaxation \cite{Wang2016}---which
is consistent with experimental findings for BZT systems \cite{Maiti2008}.
The total internal energy $E_{\textrm{int}}$ of the $H_{\textrm{eff}}$
consists of two main terms: $E_{\textrm{int}}(\{\mathrm{\mathbf{u}}_{i}\},\thinspace\{\mathbf{v}_{i}\},\thinspace\eta_{H},\thinspace\{\sigma_{j}\})=E_{\textrm{ave}}(\{\mathrm{\mathbf{u}}_{i}\},\thinspace\{\mathbf{v}_{i}\},\thinspace\eta_{H})+E_{\textrm{\ensuremath{\textrm{loc}}}}(\{\mathrm{\mathbf{u}}_{i}\},\thinspace\{\mathbf{v}_{i}\},\thinspace\{\sigma_{j}\})$,
where $\{\mathrm{\mathbf{u}}_{i}\}$ is the local soft mode in unit
cell $i$ (which is proportional to the local electric dipole moment
centered on Zr or Ti atoms), $\{\mathbf{v}_{i}\}$ are Ba-centered
local displacements related to the inhomogeneous strain inside each
cell, $\eta_{H}$ represents the homogeneous strain tensor, and $\{\sigma_{j}\}$
characterizes the $B$ sublattice atomic configuration in the BZT
solid solutions. Actually, $\sigma_{j}$ $=$ $+$1 or $-$1 corresponds
to the distribution of Zr or Ti ion located at the $j$ site of the
$B$ sublattice, respectively. The first energetic term of $E_{\textrm{ave}}$
is composed of five energetic parts: (i) the local soft mode self-energy;
(ii) the long-range dipole-dipole interaction; (iii) the short-range
interactions between local soft modes; (iv) the elastic energy; and
(v) the interaction between the local soft modes and strains \cite{Zhong1995}.
The second term, $E_{\textrm{loc}}$, represents how the distribution
of $B$ sites (Zr and Ti ions) affects the energetics involving the
local soft modes $\mathbf{u}_{i}$ and the local strain variables,
which depends on the $\{\sigma_{j}\}$ atomic configuration distribution
\cite{Akbarzadeh2012,Prosandeev2013}. In order to mimic the effect
under an applied dc electric field, an energy given by minus the dot
product between polarization and applied electric field needs to be
added to $E_{\textrm{int}}$. Note that we numerically find that the
simulated electric field is larger than the corresponding experimental
one by a factor of 100 in BZT when comparing the calculated polarization
$P$ with the experimental one for a Ba(Zr$_{0.5}$Ti$_{0.5}$)O$_{3}$
thin film at room temperature \cite{Instan2017}. Such discrepancy
is typical for atomistic effective Hamiltonian simulations \cite{Xu2017,Jiang2018}
and direct density functional theory (DFT) calculations \cite{Lu2019,Chen2019,Jiang2020,Jiang2021},
and is likely due to the fact that structural defects are not considered
in these simulations. To address such discrepancy, we presently divide
our theoretical field by a factor of 100 in the results to be reported
and discussed below. Note also that we chose here a (renormalized)
maximum applied electric field of $E_{\textrm{max}}$ $=$ $3.0\times10^{8}$
V/m, which has been experimentally achieved in Ba(Zr$_{x}$Ti$_{1-x}$)O$_{3}$
thin films \cite{Instan2017}.

Moreover, we employ the effective Hamiltonian scheme within Monte
Carlo (MC) simulations on $12\times12\times12$ supercells (8,640
atoms) with periodic boundary conditions and the distribution of Zr
and Ti ions is chosen randomly over the $B$ sublattice. In the case
of the films grown along the {[}001{]} pseudocubic direction, epitaxial
strains are associated with the freezing of some components of the
homogeneous strain tensors, namely (in Voigt notation) $\eta_{6}$
$=$ 0 and $\eta_{1}$ $=$ $\eta_{2}$ $=$ ($a_{\textrm{sub}}$$-$
$a_{\textrm{eq}}$)$/$$a_{\textrm{sub}}$, where $a_{\textrm{sub}}$
and $a_{\textrm{eq}}$ are the lattice constants of the substrate
and BZT bulk at 130 K, respectively \cite{Prosandeev2013_film}. We
also limited the simulations for strains ranging between $-$3\% and
$+$3\%, which is a physically reasonable range. Note that the films
are periodic along the three Cartesian direction and only the strain
is thus considered here when modeling epitaxial films (i.e., no surface
or thickness effect are taken into account).

\section{Results and discussion}

\subsection{Energy storage in BZT bulk}

\begin{figure}
\includegraphics[width=8cm]{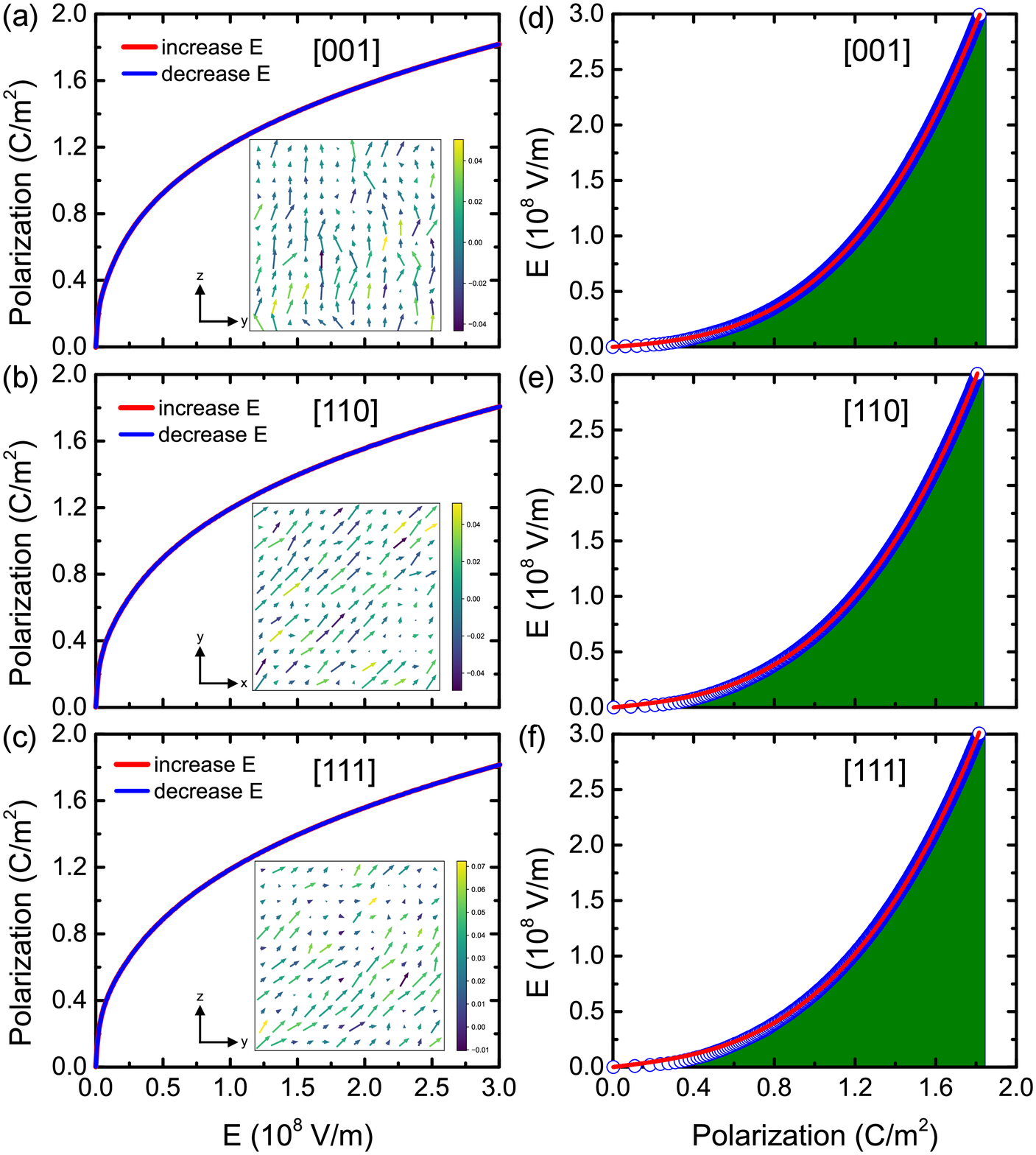}

\caption{(a)-(c) $P$-$E$ hysteresis curves at 300 K for electric field applied
along the pseudocubic {[}001{]}, {[}110{]} and {[}111{]} directions,
respectively, in BZT bulk. The insets show the dipolar configurations
in a given ($y$, $z$) or ($x$, $y$) plane at 300 K for the magnitude
of the field equal to $\simeq$$0.1\times10^{8}$ V/m with three different
field's applied directions. (d)-(f) Electric field \textit{versus}
polarization at 300 K for electric field applied along the pseudocubic
{[}001{]}, {[}110{]} and {[}111{]} directions, respectively. The green
areas represent the energy densities and the solid red lines represent
the fit of the MC data by the Landau model. \label{fig:P-E_bulk}}
\end{figure}

Figures~\ref{fig:P-E_bulk}(a)-\ref{fig:P-E_bulk}(c) show the $P$-$E$
curves at a selected room temperature of 300 K for electric field
applied along the pseudocubic {[}001{]}, {[}110{]} and {[}111{]} directions,
respectively, in BZT bulk. For all these directions, the resulting
polarization aligns along the field. As depicted in Figs.~\ref{fig:P-E_bulk}(a)-\ref{fig:P-E_bulk}(c),
the charging and discharging processes of the $P$-$E$ curves are
completely reversible, which implies that the energy efficiency is
100\% (the charging and discharging cycles correspond to electric
field increasing from zero to $E_{\textrm{max}}$ $=$ $3.0\times10^{8}$
V/m and then decreasing back to zero field, respectively). Note that
the ideal 100\% efficiency has also been predicted in epitaxial AlN/ScN
superlattices \cite{Jiang2021} as a result of a field-induced \textit{second-order}
transition towards a ferroelectric state. In the case of BZT, measurements
also found a large value for this efficiency, namely of 82.46$\pm$1.0\%
at an electric field $\simeq$3.0$\times$10$^{8}$ V/m \cite{Instan2017}
(the deviation from the ideal value of 100\% may also arise from the
presence of defects in the grown sample).

Moreover, the insets of Figs.~\ref{fig:P-E_bulk}(a)-\ref{fig:P-E_bulk}(c)
display the local dipole configurations in a given ($y$, $z$) or
($x$, $y$) plane at 300 K for these three different applied field
directions with a magnitude of $\simeq$$0.1\times10^{8}$ V/m. Recalling
that the $H_{\textrm{eff}}$ predicted that, under zero field and
for temperatures below the Burns temperature ($T_{\textrm{b}}$ $\simeq$
450 K) \cite{Akbarzadeh2012}, BZT possesses different polar nanoregions
for which the dipoles align along different $\left\langle 111\right\rangle $
pseudocubic directions (hence resulting in an overall vanishing polarization),
these insets demonstrate that the application of electric field along
a given direction forces dipoles to rotate towards the field's direction
(hence giving rise to a finite polarization along that latter direction).

Furthermore, Figs.~\ref{fig:P-E_bulk}(d)-\ref{fig:P-E_bulk}(f)
display the electric field applied along these {[}001{]}, {[}110{]}
and {[}111{]} pseudocubic directions as a function of polarization
for temperature at 300 K, which allows us to extract the energy density
since it is simply the green area shown in these figures.

\begin{figure}
\includegraphics[width=8cm]{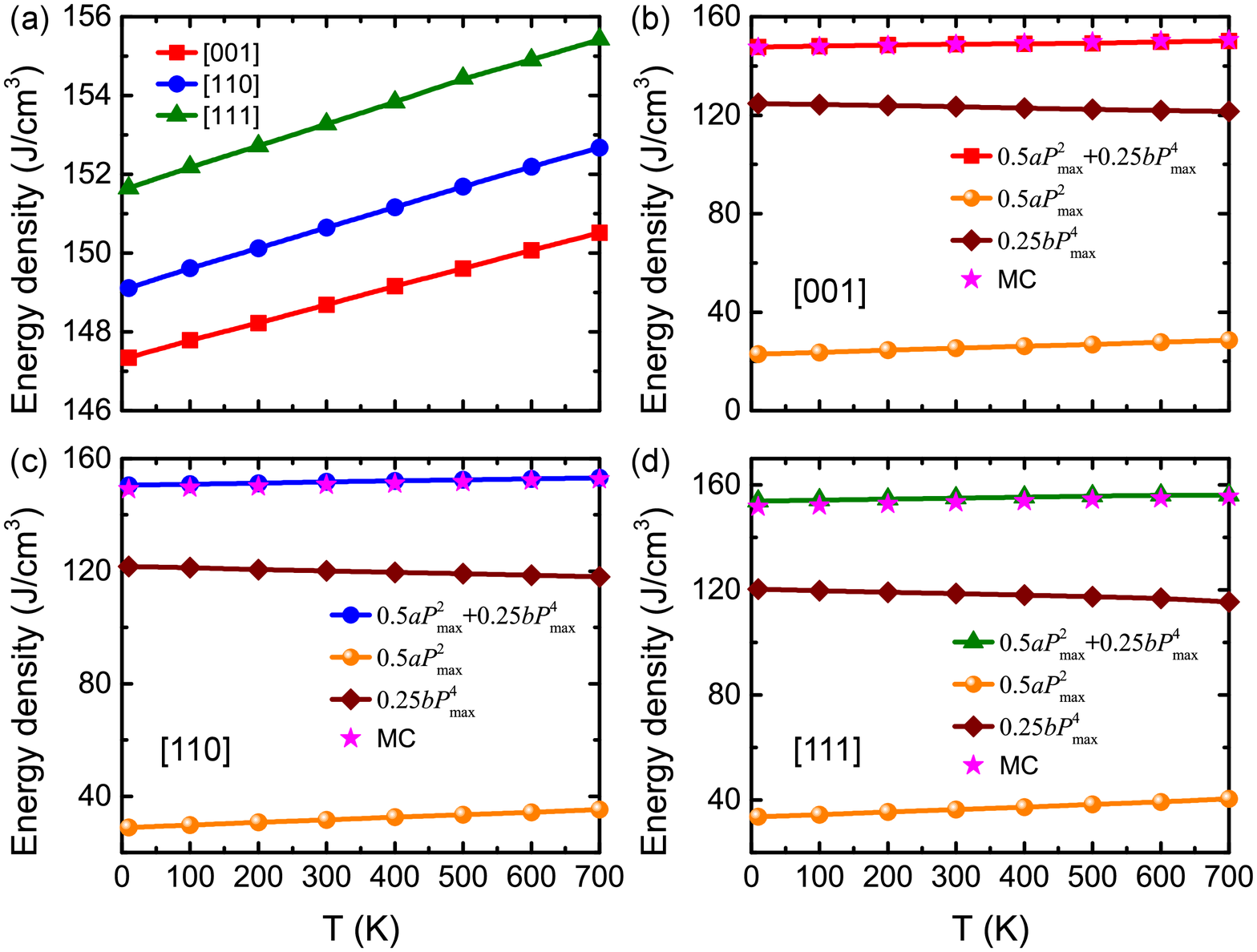}

\caption{(a) Energy density obtained from MC data as a function of temperature
for electric field applied along the {[}001{]}, {[}110{]} and {[}111{]}
directions, with a maximal electric field being equal to $3.0\times10^{8}$
V/m, in BZT bulk. (b)-(d) Total and decomposed energy densities obtained
from Eq.~(\ref{eq:energy density Landau}) as a function of temperature
at $E_{\textrm{max}}$ $=$ $3.0\times10^{8}$ V/m for electric field
applied along the {[}001{]}, {[}110{]} and {[}111{]} directions, respectively.
Stars also display the MC data of the total energy densities again
in Panels (b)-(d) for comparison. \label{fig:enegy_density_bulk}}
\end{figure}

Doing that for all considered temperatures (up to 700 K) therefore
allows us to compute the energy densities as a function of temperature
for electric field applied along the {[}001{]}, {[}110{]} and {[}111{]}
pseudocubic directions, which are shown in Fig.~\ref{fig:enegy_density_bulk}(a).
It is interesting to realize that (1) the energy densities linearly
increase with temperature for the three different considered field's
directions; and (2) we predict ultrahigh energy densities and values
varying between 147 and 155 J/cm$^{3}$, which agrees very well with
experimental reports in Ba(Zr$_{x}$Ti$_{1-x}$)O$_{3}$ thin films
\cite{Instan2017}.

In order to understand the origin of these energy density features,
let us take advantage of the simple Landau-type free energy model
developed in Ref.~\cite{Jiang2021}:

\begin{equation}
F=\frac{1}{2}aP^{2}+\frac{1}{4}bP^{4}-EP,\label{eq:Landau}
\end{equation}
where $a$ and $b$ are quadratic and quartic coefficients, respectively.

At equilibrium, one must have $\frac{\partial F}{\partial P}=0$,
which thus leads to:

\begin{equation}
E=aP+bP^{3}.\label{eq:Landau-1}
\end{equation}

\begin{figure}
\includegraphics[width=8cm]{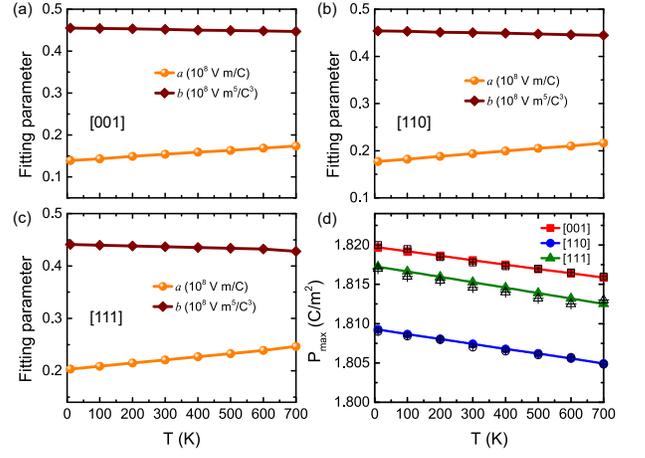}

\caption{(a)-(c) Temperature dependence of fitting parameters $a$ and $b$
(see text) for electric field applied along the {[}001{]}, {[}110{]}
and {[}111{]} directions, respectively, with a maximal applied electric
field of $E_{\textrm{max}}$ $=$ $3.0\times10^{8}$ V/m, in BZT bulk.
(d) $P_{\textrm{max}}$ obtained from MC data (filled symbols) and
Eq.~(\ref{eq:Landau-1}) (open symbols) as a function of temperature
at $E_{\textrm{max}}$ $=$ $3.0\times10^{8}$ V/m for electric field
applied along the {[}001{]}, {[}110{]} and {[}111{]} directions, in
BZT bulk. \label{fig:fitting_parameters_Pmax}}
\end{figure}

Interestingly, the electric field \textit{versus} polarization ($E$-$P$)
MC data for all considered temperatures can indeed be nicely fitted
by Eq.~(\ref{eq:Landau-1}), which shows that such equation is valid
but also allows to extract the $a$ and $b$ parameters for each temperature
and considered field's direction. These latter $a$ and $b$ coefficients
are shown in Figs.~\ref{fig:fitting_parameters_Pmax}(a)-\ref{fig:fitting_parameters_Pmax}(c)
as a function of temperature for fields up to the maximum field $E_{\textrm{max}}$
$=$ $3.0\times10^{8}$ V/m applied along the {[}001{]}, {[}110{]}
and {[}111{]} pseudocubic directions, respectively. These $a$ and
$b$ coefficients are important for energy storage since they are
involved in the expression of the energy density, according to the
Landau model \cite{Jiang2021}:

\begin{equation}
U=\int_{0}^{P_{\textrm{max}}}(aP+bP^{3})dP=\frac{1}{2}aP_{\textrm{max}}^{2}+\frac{1}{4}bP_{\textrm{max}}^{4},\label{eq:energy density Landau}
\end{equation}
where $P_{\textrm{max}}$ is the polarization at $E_{\textrm{max}}$.
Note that, practically, one can determine $P_{\textrm{max}}$ via
two different ways, that are directly from the MC data or via Eq.~(\ref{eq:Landau-1})
at the field of $E_{\textrm{max}}$ since $a$ and $b$ parameters
are now known. As shown in Fig.~\ref{fig:fitting_parameters_Pmax}(d),
these two methods give nearly identical results.

Equation (\ref{eq:energy density Landau}) therefore tells us that
only three quantities completely govern the behaviors and values of
the energy density, namely $a$, $b$ and $P_{\textrm{max}}$. It
is thus necessary to comment on their behaviors with temperature and
field's direction that are reported in Fig.~\ref{fig:fitting_parameters_Pmax}.
The fitting parameter $a$ linearly increases with temperature for
the three different field's directions---as expected from usual Landau
theory of ferroelectrics stating that this coefficient should be equal
to $a_{0}(T-T_{c})$, where $a_{0}$ is a positive constant and $T_{c}$
is a critical temperature (that can be negative for relaxor ferroelectrics)
\cite{Chandra2007}. It varies from from 0.139$\times$10$^{8}$ to
0.174$\times$10$^{8}$ V m/C for the {[}001{]} direction, from 0.177$\times$10$^{8}$
to 0.216$\times$10$^{8}$ V m/C for {[}110{]}, and from 0.203$\times$10$^{8}$
to 0.247$\times$10$^{8}$ V m/C for {[}111{]}, respectively, for
temperatures ranging between 10 K and 700 K. In contrast, the $b$
parameter only slightly linearly decreases with temperature (its value
concomitantly ranges from 0.455$\times$10$^{8}$ to 0.447$\times$10$^{8}$
V m$^{5}$/C$^{3}$ for {[}001{]}, from 0.454$\times$10$^{8}$ to
0.445$\times$10$^{8}$ V m$^{5}$/C$^{3}$ for {[}110{]}, and from
0.441$\times$10$^{8}$ to 0.428$\times$10$^{8}$ V m$^{5}$/C$^{3}$
for {[}111{]}, respectively). Finally, $P_{\textrm{max}}$ basically
only very slightly linearly decreases with temperature from 1.820
to 1.816 C/m$^{\textrm{2}}$ for fields applied along {[}001{]}, from
1.809 to 1.805 C/m$^{\textrm{2}}$ for {[}110{]}, and from 1.817 to
1.813 C/m$^{\textrm{2}}$ for {[}111{]}, respectively, with temperatures
varying from 10 K to 700 K. Its rather large value indicates that
BZT is easily polarizable. The parameter that is the most sensitive
to both temperature and field's direction is therefore the $a$ coefficient.
Its temperature behavior (linear increase and always positive values)
basically indicates that increasing the temperature makes BZT going
further away from a ferroelectric state at zero field for any direction
of the field. Its dependency on field's direction at any temperature
(smaller positive values $a$ for the {[}001{]} direction and larger
values for {[}111{]}) reveals that it is easier to induce a ferroelectric
state when applying an electric field along {[}001{]} than {[}110{]}
and then {[}111{]}---likely because inducing a ferroelectric state
with a polarization along {[}111{]} requires the polar nanoregions
existing at zero field and having electric dipoles along {[}$\bar{1}$$\bar{1}$$\bar{1}${]}
to completely revert their polarization rather than simply rotate
towards an intermediate direction, such as {[}001{]}.

The behaviors of $a$, $b$ and $P_{\textrm{max}}$ allow us to understand
the results of the energy density in Fig.~\ref{fig:enegy_density_bulk}(a)
since Eq.~(\ref{eq:energy density Landau}) indicates that such energy
density can be decomposed in two terms, that are $\frac{1}{2}aP_{\textrm{max}}^{2}$
and $\frac{1}{4}bP_{\textrm{max}}^{4}$, and which are shown in Figs.~\ref{fig:enegy_density_bulk}(b)-\ref{fig:enegy_density_bulk}(d)
for the three different field's direction with $E_{\textrm{max}}$
$=$ $3.0\times10^{8}$ V/m, along with the (total) energy densities
directly obtained from the MC data. One can clearly see that, for
any considered temperature, Eq.~(\ref{eq:energy density Landau})
and the MC energy densities provide nearly identical results. Since
$b$ and $P_{\textrm{max}}$ are in first approximation independent
of both the temperature and the field's direction, the dependencies
of the total energy density on temperature and crystallographic direction
of the electric field basically arise from the aforementioned corresponding
dependencies of the $a$ parameter. Note that the contribution of
$\frac{1}{2}aP_{\textrm{max}}^{2}$ ($\frac{1}{4}bP_{\textrm{max}}^{4}$)
to the total energy density at 300 K is 17\% (83\%), 21\% (79\%),
and 23\% (77\%) for electric field applied along the {[}001{]}, {[}110{]}
and {[}111{]} directions, respectively. These numbers, as well as
Fig.~\ref{fig:enegy_density_bulk}, Fig.~\ref{fig:fitting_parameters_Pmax}
and Eq.~(\ref{eq:energy density Landau}), therefore tell us that
having a large energy density accompanied by a large efficiency can
be accomplished by (1) having large positive $a$ and $b$ coefficients,
which characterize systems that can not be energetically too close
to a ferroelectric state and that will undergo a second-order field-induced
transition to a ferroelectric state at large fields; while (2) having
a large $P_{\textrm{max}}$ at feasible electric fields, which indicates
that the system is easily polarizable at these fields and which thus
also implies that $a$ and $b$ can not be too large. Note that conditions
(1) and (2) are relevant to analyze and understand the ultrahigh energy
storage in initially non-polar AlN/ScN superlattices \cite{Jiang2021}
and superparaelectric relaxor ferroelectrics \cite{Pan2021}.

\subsection{Energy storage in (001) BZT films}

\begin{figure}
\includegraphics[width=8cm]{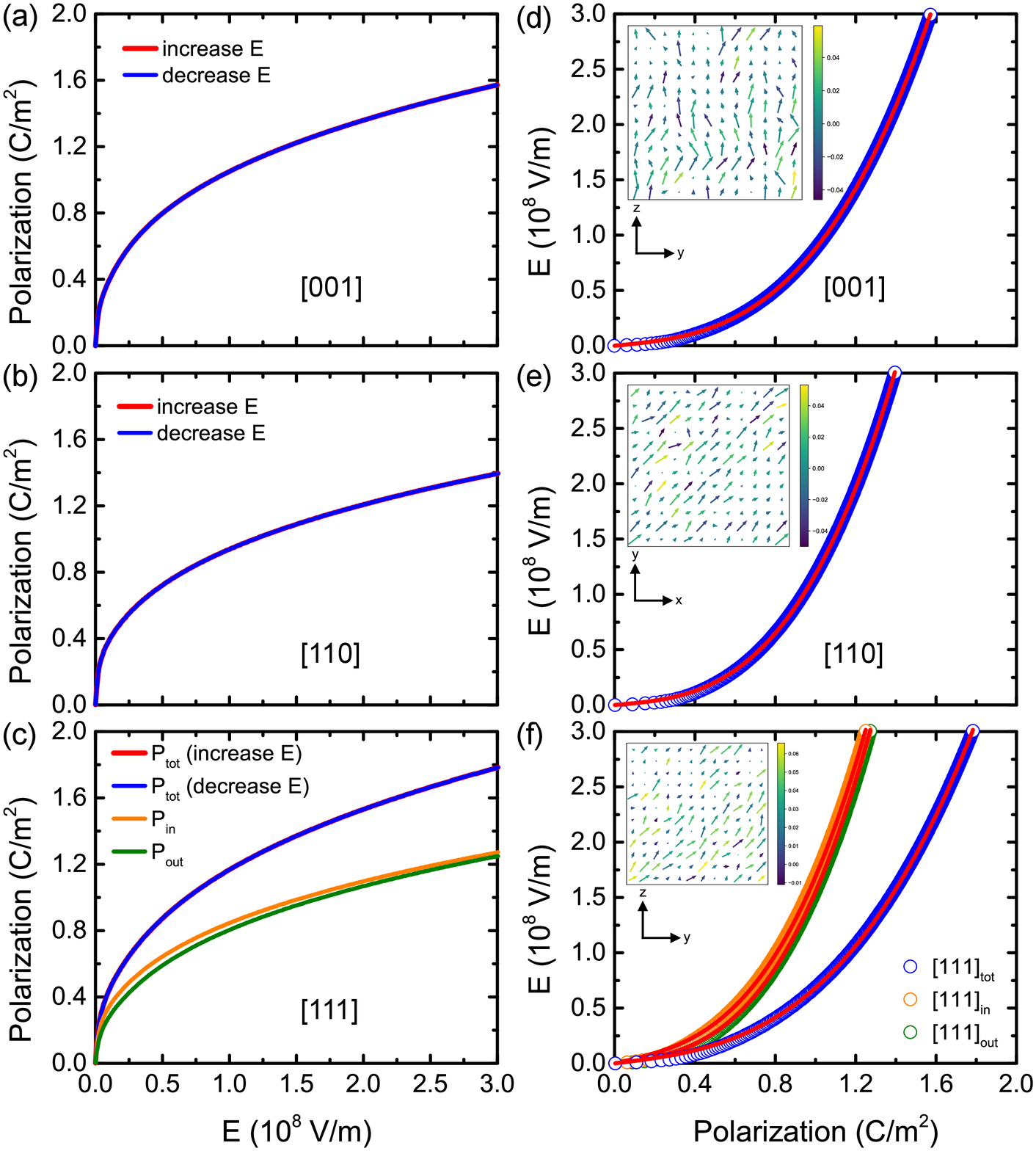}

\caption{Same as Fig.~\ref{fig:P-E_bulk} but for (001) BZT films at 0\% strain.
\label{fig:P-E_film}}
\end{figure}

Let us now present the energy storage results for (001) BZT films
as a function of epitaxial strain for the selected temperature of
300 K. Figures~\ref{fig:P-E_film}(a)-\ref{fig:P-E_film}(c) respectively
show the $P$-$E$ curves at 300 K and zero strain for field applied
along (i) the pseudocubic {[}001{]} that results in a polarization
lying along that out-of-plane direction; (ii) the {[}110{]} direction
that yields a polarization aligned along that in-plane direction;
and (iii) along the {[}111{]} direction that induces a polarization
being along $[uuv]$ directions, that is $P_{x}$$=$ $P_{y}$ $\neq$
$P_{z}$, due to the freezing of $\eta_{1}$ $=$ $\eta_{2}$ strain
components while $\eta_{3}$ can relax. Note that the direction of
these polarization in the three cases is also consistent with the
insets of Figs.~\ref{fig:P-E_film}(d)-\ref{fig:P-E_film}(f) showing
the dipole configurations in a given ($y$, $z$) or ($x$, $y$)
plane at zero strain and 300 K. Note also that, consequently, we show
three types of $P$-$E$ data for fields applied along the {[}111{]}
direction in Fig.~\ref{fig:P-E_film}(c), that is for the in-plane
component of the polarization $P_{\textrm{in}}$ (which is along the
{[}110{]} direction), the out-of-plane component of the polarization
$P_{\textrm{out}}$ (that is along {[}001{]}) and the total polarization
$P_{\textrm{tot}}$ (that is given by $P_{\textrm{tot}}$ $=$ $\sqrt{P_{x}^{2}+P_{y}^{2}+P_{z}^{2}}$).

\begin{figure}
\includegraphics[width=7cm]{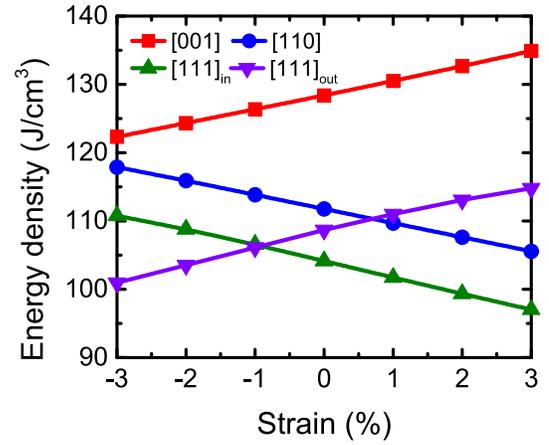}

\caption{Energy density obtained from MC data \textit{versus} strain at 300
K and $E_{\textrm{max}}$ $=$ $3.0\times10^{8}$ V/m for fields applied
along the {[}001{]}, {[}110{]} and {[}111{]} directions, respectively,
in (001) BZT films. The distinction between {[}111{]}$_{\textrm{in}}$
\textit{versus} {[}111{]}$_{\textrm{out}}$ is explained in the text,
in case of fields applied along {[}111{]}. \label{fig:enegy_density_film_MC}}
\end{figure}

Moreover, Figs.~\ref{fig:P-E_film}(d)-\ref{fig:P-E_film}(f) show
the corresponding $E$-$P$ data at zero strain and 300 K for fields
applied along these three different directions up to $E_{\textrm{max}}$
$=$ $3.0\times10^{8}$ V/m, which, once again, allows the energy
density to be extracted via the computation of areas similar to the
green one of Figs.~\ref{fig:P-E_bulk}(d)-\ref{fig:P-E_bulk}(f).
Such types of calculations are then performed for all considered strains
at 300 K, which yields the results for energy density reported in
Fig.~\ref{fig:enegy_density_film_MC} for the different directions
of the field. Note that we continue to distinguish between in-plane
\textit{versus} out-of-plane components of the polarization in case
of a field applied along {[}111{]}, with this distinction resulting
in the wording of {[}111{]}$_{\textrm{in}}$ \textit{versus} {[}111{]}$_{\textrm{out}}$
in Fig.~\ref{fig:P-E_film}(f) and Fig.~\ref{fig:enegy_density_film_MC}
as well as in the text.

One can first realize that for all considered strains, the energy
density is still large in magnitude in the films (typically larger
than 100 J/cm$^{3}$), while being smaller than that of BZT bulk at
room temperature. For instance, the energy densities of bulk BZT at
300 K are 148.7 J/cm$^{3}$ when the field is applied along {[}001{]},
150.6 J/cm$^{3}$ for {[}110{]}, and 153.3 J/cm$^{3}$ for {[}111{]},
respectively. This slight decrease when going from bulk to films originates
from the fact that, for each considered field and at a fixed temperature,
the in-plane lattice constants are constant in the epitaxial films
while they can vary when changing the fields in the bulk---which
can thus give rise to larger polarizations in bulk than in films.
Other striking features of Fig.~\ref{fig:enegy_density_film_MC},
are that the energy densities resulting from the application of a
field along {[}001{]} as well as the one associated with the out-of-plane
component of polarization under a {[}111{]} field, that is {[}111{]}$_{\textrm{out}}$,
both linearly increase with strain ranging between $-$3\% and $+$3\%.
In contrast, the energy densities when the field is applied along
{[}110{]} and the other one linked with in-plane component of polarization,
namely {[}111{]}$_{\textrm{in}}$, both linearly decrease with such
strain. Furthermore, the largest energy density is found for our maximal
considered tensile strain (that is $+$3\%) in case of a field applied
along {[}001{]}.

\begin{figure}
\includegraphics[width=8cm]{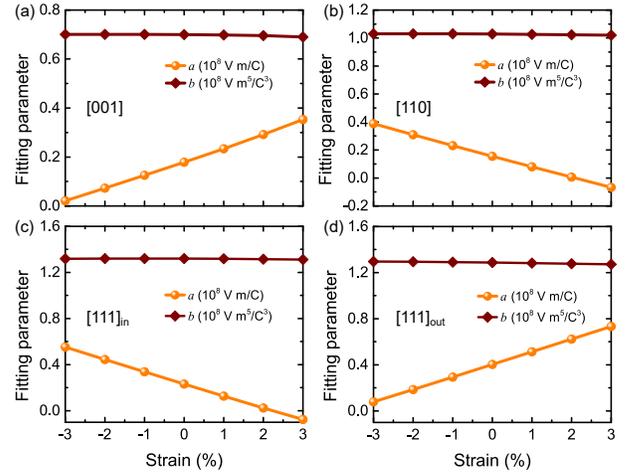}

\caption{Panels (a) and (b): Same as Figs.~\ref{fig:fitting_parameters_Pmax}(a)
and \ref{fig:fitting_parameters_Pmax}(b) but as a function of strain
at 300 K in BZT films. Panels (c) and (d) show the in-plane and out-of-plane
fitting parameters $a$ and $b$ when field is applied along the {[}111{]}
direction, respectively. \label{fig:fitting_parameters_film}}
\end{figure}

\begin{figure}
\includegraphics[width=7cm]{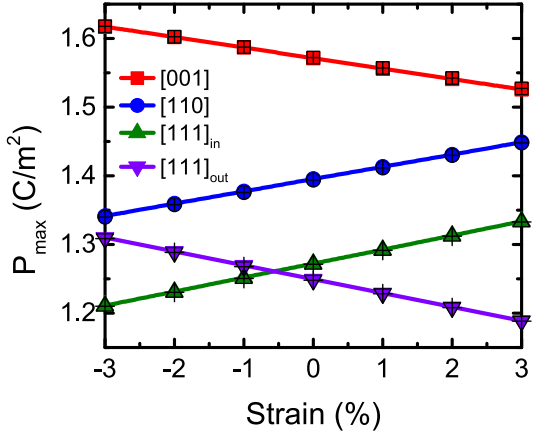}

\caption{Same as Fig.~\ref{fig:fitting_parameters_Pmax}(d) but as a function
of strain at 300 K in BZT films. \label{fig:Pmax_films}}
\end{figure}

To understand the energy density results in Fig.~\ref{fig:enegy_density_film_MC}
for (001) BZT films, we also decided to use Eq.~(\ref{eq:Landau-1})
(which is once again found to fit well the MC data) and Eq.~(\ref{eq:energy density Landau}).
Consequently, Fig.~\ref{fig:fitting_parameters_film} shows the $a$
and $b$ fitting parameters while Fig.~\ref{fig:Pmax_films} displays
the resulting $P_{\textrm{max}}$, for these films. Figure~\ref{fig:fitting_parameters_film}(a)
and Figure~\ref{fig:Pmax_films} indicate that, when the field is
applied along the {[}001{]} direction, the (i) fitting parameter $a$
is positive and linearly increases when the strain increases from
$-$3\% to $+$3\% ($a$ varies from 0.02$\times$10$^{8}$ to 0.35$\times$10$^{8}$
V m/C); (ii) the $b$ coefficient is basically a constant with strain
(it varies from 0.70$\times$10$^{8}$ to 0.69$\times$10$^{8}$ V
m$^{5}$/C$^{3}$); and (iii) $P_{\textrm{max}}$ linearly decreases
with strain from 1.62 to 1.53 C/m$^{\textrm{2}}$. In contrast, when
the field is along {[}110{]}, Fig.~\ref{fig:fitting_parameters_film}(b)
and Fig.~\ref{fig:Pmax_films} reveal that (iv) $a$ linearly decreases
with strain (from 0.39$\times$10$^{8}$ to $-$0.07$\times$10$^{8}$
V m/C, therefore becoming slightly negative at large tensile strains);
(v) $P_{\textrm{max}}$ linearly increases with strain (from 1.34
to 1.45 C/m$^{\textrm{2}}$); (vi) while $b$ continues to be basically
constant with strain (it only changes from 1.03$\times$10$^{8}$
to 1.02$\times$10$^{8}$ V m$^{5}$/C$^{3}$). Items (i)-(vi) can
be simply understood by realizing that increasing strain from compressive
to tensile in epitaxial (001) films is known to progressively disfavor
a ferroelectric state with an out-of-plane polarization in favor of
a ferroelectric state with an in-plane polarization \cite{Dupe2011,Yang2012,Chen2015}.
For the same reasons and as shown in Fig.~\ref{fig:fitting_parameters_film}(c),
Fig.~\ref{fig:fitting_parameters_film}(d) and Fig.~\ref{fig:Pmax_films},
the $a$ parameter and $P_{\textrm{max}}$ associated with the out-of-plane
(respectively, in-plane) components for fields applied along {[}111{]}
have the same behavior with strain as those for the field lying along
{[}001{]} (respectively, {[}110{]}). Note also that the $b$ parameter
continues to be basically independent of strain in the case of fields
applied {[}111{]}, as well---which reveals that BZT can adopt \textit{second-order}
phase transition when under strain and/or field \cite{Prosandeev2013}.

\begin{figure}
\includegraphics[width=8cm]{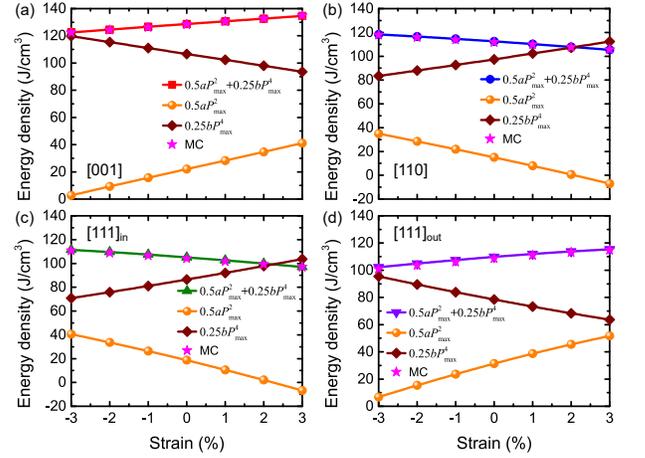}

\caption{Panels (a)-(d): Total and decomposed energy densities \textit{versus}
strain for field applied along {[}001{]} {[}Panel (a){]}, {[}110{]}
{[}Panel (b){]} and {[}111{]} {[}Panels (c) and (d) for in-plane \textit{versus}
out-of-plane components of the polarization{]}, at a maximal applied
electric field equal to $3.0\times10^{8}$ V/m and 300 K in BZT films.
\label{fig:energy_density_landau_film}}
\end{figure}

Let us now pay attention to the two terms of Eq.~(\ref{eq:energy density Landau})
that sump up to be the total energy density. They are shown in Fig.~\ref{fig:energy_density_landau_film},
using the fitting parameters ($a$ and $b$) of Fig.~\ref{fig:fitting_parameters_film}
and $P_{\textrm{max}}$ of Fig.~\ref{fig:Pmax_films}. Due to the
aforementioned behavior of $a$ that grows faster than $P_{\textrm{max}}$
decreases with strain, the first contribution, $\frac{1}{2}aP_{\textrm{max}}^{2}$,
increases with strain (from 2.7 for $-$3\% to 41.1 J/cm$^{\textrm{3}}$
for $+$3\%), when the field is applied along the {[}001{]} direction---as
displayed in Fig.~\ref{fig:energy_density_landau_film}(a). In contrast,
the second contribution, $\frac{1}{4}bP_{\textrm{max}}^{4}$, decreases
with strain from 119.8 to 93.6 J/cm$^{\textrm{3}}$) when the field
is also along this {[}001{]} direction, as a result of the concomitant
decrease of $P_{\textrm{max}}$ while $b$ is basically constant with
strain. Note that the change of value in $\frac{1}{4}bP_{\textrm{max}}^{4}$
is smaller than $\frac{1}{2}aP_{\textrm{max}}^{2}$ for this field's
direction, once again because of the fast increase of $a$ with strain
reflecting the desire of the system to be energetically far from a
polarized state with a polarization along {[}001{]} when enhancing
the strain from compressive to tensile values. Consequently, the total
energy density increases with strain.

The behaviors are opposite when the field is applied along {[}110{]}
because BZT films become much closer in energy to adopt a ferroelectric
state with an in-plane {[}110{]} polarization direction as the strain
is enhanced. Consequently, $a$ strongly decreases with strain while
$P_{\textrm{max}}$ is enhanced but at a smaller extent. As a result,
$\frac{1}{2}aP_{\textrm{max}}^{2}$ decreases with strain (values
varying between 34.9 and $-$7.1 J/cm$^{3}$) faster than $\frac{1}{4}bP_{\textrm{max}}^{4}$
increases (from 83.5 to 112.4 J/cm$^{\textrm{3}}$) for field applied
along the {[}110{]} direction---as shown in Fig.~\ref{fig:energy_density_landau_film}(b).
The resulting total energy density thus decreases with strain.

Regarding the energy density for fields applied along the {[}111{]}
direction, Fig.~\ref{fig:energy_density_landau_film}(c) {[}respectively,
Fig.~\ref{fig:energy_density_landau_film}(d){]} that is related
to the in-plane (respectively, out-of-plane) component of the polarization
shows that the behaviors of total and decomposed energy densities
are very similar to Fig.~\ref{fig:energy_density_landau_film}(b)
that corresponds the field applied along {[}110{]} (respectively to
Fig.~\ref{fig:energy_density_landau_film}(a) that corresponds the
field applied along {[}001{]}) for the same energetic reasons, i.e.,
going from compressive strain to tensile strain favors the formation
of in-plane polarization while disfavoring ferroelectric states with
out-of-plane polarization.

\section{Summary}

In summary, an atomistic effective Hamiltonian scheme combined with
Monte-Carlo simulations was used to investigate the energy storage
in bulk and epitaxial (001) films made of BZT. We find that these
BZT systems can exhibit ultrahigh energy densities and an ideal efficiency
of 100\%. These energy storage results are then interpreted via a
simple phenomenological model that reproduces these MC data. More
precisely, energy density can be decomposed in two terms: the first
term being the product of the fitting $a$ parameter and $P_{\textrm{max}}^{2}$,
and the second term being the product between the $b$ parameter and
$P_{\textrm{max}}^{4}$. The behavior of $a$, $b$ and $P_{\textrm{max}}$
lead to a competition between these two terms (that can be understood
in terms of energetics) that eventually causes the temperature-dependency,
field's direction dependency and strain-dependency of the total energy
density. The proposed phenomenological model can be easily employed
for nonlinear dielectrics with large energy density. We thus hope
that the present article deepens the fields of energy storage in relaxor
ferroelectrics and other nonlinear dielectrics. 
\begin{acknowledgments}
This work is supported by the National Natural Science Foundation
of China (Grant No.\ 11804138), Shandong Provincial Natural Science
Foundation (Grant No.\ ZR2019QA008), China Postdoctoral Science Foundation
(Grants No.\ 2020T130120 and No.\ 2018M641905), and ``Young Talent
Support Plan'' of Xi'an Jiaotong University (Grant No.\ WL6J004).
S.P and L.B.\ acknowledge the Office of Naval Research for the support
under Grants No.\ N00014-17-1-2818 and No.\ N00014-21-1-2086. L.B.\ also
acknowledges ARO Grant No.\ W911NF-21-1-0113 and DARPA Grant No.\ HR0011-15-2-0038
(MATRIX program). The HPC Platform of Xi'an Jiaotong University is
also acknowledged. 
\end{acknowledgments}

\end{document}